\documentclass[notitlepage,aip,reprint,onecolumn]{revtex4-1}
\usepackage{graphicx}
\usepackage{physics}
\usepackage{amsmath}
\usepackage{booktabs}
\usepackage{siunitx}
\usepackage{orcidlink}

\begin{document}

\title{Limits on the atomic description of four-wave mixing}

\author{Irvin F. Ángeles-Aguillón \orcidlink{0000-0002-1383-5037}}
\author{Nieves Arias-Téllez \orcidlink{0000-0000-0000-0000}}
\author{Pablo Yanes-Thomas \orcidlink{0009-0001-8047-9853}}
\affiliation{Departamento de Física Cuántica y Fotónica, Instituto de Física,
Circuito Exterior, Universidad Nacional Autónoma de México, Ciudad de México,
CP 04510, México}
\author{Alejandro Kunold\orcidlink{0000-0000-0000-0000}}
\email{akb@azc.uam.mx}
\affiliation{Área de Física Teórica y Materia Condensada,
Universidad Autónoma Metropolitana Azcapotzalco,
Ciudad de México 02200, Mexico}
\author{Daniel Sahagún-Sánchez\orcidlink{0000-0001-8382-9727}}
\email{sahagun@fisica.unam.mx }
\affiliation{Departamento de Física Cuántica y Fotónica, Instituto de Física,
Circuito Exterior, Universidad Nacional Autónoma de México, Ciudad de México,
CP 04510, México}

\keywords{four-wave mixing, atomic gases, quantum information}

\begin{abstract}
The limits of the well-established single-atom model for describing photon-pair
generation via four-wave mixing in a diamond configuration in atomic
ensembles are experimentally tested.
Using a cold-atom source, biphotons are generated and detected through
polarization analyzers that resolve the emitted light into horizontal
and vertical components in the laboratory frame.
The pump lasers driving the first and second excitation transitions
are horizontally and vertically polarized, respectively.
To predict single-photon counts and coincidence rates in all
polarization channels, the first- and second-order correlation
functions are calculated directly using a comprehensive model that
accounts for all Zeeman sublevels of the relevant hyperfine states.
Furthermore, the population dynamics with and without the re-pump
laser of the magneto-optical trap are compared, revealing substantial
differences in the populations of the Zeeman sublevels.
This motivates the inclusion of two additional hyperfine levels and
their corresponding Zeeman sublevels in the final model.
The resulting density matrix is used to calculate expectation values
of the far-field electric-field operators and compare them with
experimental measurements over a range of pump-laser powers.
Excellent agreement with the measured photon counts is obtained over
most of this range.
For the coincidence measurements, good agreement is found when the
polarization of the photon generated by the first (second) decay is
parallel to that of the first (second) pump beam.
In contrast, the model consistently underestimates the experimental
coincidence rates for the opposite polarization configuration.
These findings indicate that effects beyond the internal level
dynamics of individual atoms, most notably collective phenomena, are
required to fully account for photon coincidences generated by
four-wave mixing in atomic ensembles.
\end{abstract}

\maketitle

\section{Introduction}

Atomic systems are a leading physical platform for key
tasks in quantum technology such as information processing~\cite{Menon2024,Edlbauer2025}
and storage~\cite{Ma2022,Noh2021,Wu2022}.
Pursued applications in telecommunications~\cite{Chaneliere:2006gx}
and in the quantum internet itself~\cite{Kimble:2008if,wehner2018,azuma2023}
necessitate the ability to encode quantum information in single photons,
allowing it to be transmitted between single atoms or atomic ensembles.
In these applications, photons constitute ideal flying qubits for
transmitting quantum information between spatially separated quantum
nodes~\cite{Bouwmeester1997}.

Probabilistic and on-demand photon sources can be realized using
nonlinear crystals
\cite{Du2008nonlinear,Wolfgramm2011,Srivathsan2013,Farrera2016,
Rambach2016,slattery2019cavity,Prakash2019},
optical fibers \cite{garaypalmett2023},
quantum dots \cite{Akopian2006,Alqedra2025,Zhu2025},
color centers in diamond \cite{Cheng2013,Rao2015,Shan2018},
ions \cite{Simon2003,Maunz2007},
and neutral atoms \cite{Eisaman:2011}.
Atomic systems offer the potential advantage of accurately and
flexibly tailoring photonic qubits, since the degrees of freedom of
the emitted light are intimately related to the internal atomic
states, which can nowadays be exquisitely controlled using techniques
developed in precision spectroscopy.
Additionally, through the nonlinear process of four-wave mixing (FWM),
atomic ensembles can naturally generate photons with bandwidths in the
few-megahertz range \cite{du2008,shu2019,Srivathsan:2013fa}.
Such narrow bandwidths are challenging to achieve with other photon-source
platforms but are essential for quantum networks based on atomic
nodes \cite{covey2023}.

The advantages offered by atomic FWM have motivated extensive
experimental and theoretical efforts to understand and optimize the
underlying photon-generation process. Despite these advances, however,
the internal atomic dynamics are still commonly described using highly
simplified models.
Thus far, most theoretical descriptions of FWM in alkali atoms rely on
few-level models that drastically reduce the internal electronic
structure by neglecting the full hyperfine and Zeeman manifolds
\cite{Willis2009,Khadka2012,Lee2017,Leszczynski2017,Cere:2018ds,Gladysz2024}.
Consequently, although these approaches capture interference phenomena
between the atomic levels retained in the model, they generally fail
to (i) incorporate established protocols for atomic-sample preparation,
such as optical pumping and re-pumping, and (ii) fully relate the
internal atomic dynamics to the polarization of the generated photons.
Such simplified models have nevertheless provided sufficient predictive
power to support numerous proof-of-principle demonstrations.
However, their limitations become apparent when attempting a detailed
quantitative characterization of photon-pair sources.
Indeed, even for cold-atom systems, where complications such as thermal
motion are strongly suppressed, few-level models have been shown to
provide an incomplete description of the observed photon statistics
\cite{Cere:2018ds}. 

Extending the single-atom model to incorporate many-body phenomena
is technically and computationally challenging
\cite{Lehmberg:1970,Yanes-Thomas:2025,Yanes-Thomas:2026}.
It is therefore important to establish the limits of the
single-atom description by determining which features of the internal
atomic structure are essential for reproducing the observed dynamics.
This is the objective of the present work.

Specifically, we investigate the power dependence of photon counts
and coincidence rates generated by inducing FWM within a diamond scheme
\cite{Willis2008} in a cold atomic ensemble confined in a
magneto-optical trap.
To isolate the role of the internal atomic structure, we develop a
comprehensive single-atom 32-level model that explicitly includes all
relevant hyperfine and Zeeman sublevels, together with the re-pump
process, thereby extending the simplified few-level models commonly
employed in the literature.
Unlike previous approaches, which often require separate calculations
or independent parameter fits for different polarization channels, the
present model describes all relevant optical transitions and
polarization-resolved emissions within a single, self-consistent
framework using a common set of physical parameters.
The model is formulated within the Lindblad master-equation framework
\cite{Lehmberg:1970,Genes2022,Yanes-Thomas:2025,Yanes-Thomas:2026},
from which the far-field electric field and the first- and second-order
correlation functions are calculated directly.

We show that the proposed model provides good quantitative agreement
with the measured first-order correlation functions, demonstrating that
an accurate description of the internal atomic structure is essential
for reproducing the observed photon counts.
For the second-order correlation functions, good agreement is also
achieved when the polarization of each generated photon is parallel
to that associated with its corresponding decay transition.
In contrast, for the opposite polarization configuration, the model
consistently underestimates the experimental coincidence rates.
Since the present treatment neglects interatomic correlations, this
remaining discrepancy suggests that collective effects, rather than an
incomplete description of the internal atomic structure, play an
important role in the observed coincidence measurements.

The remainder of this paper is organized as follows.
In Section~\ref{sec:system}, we introduce the particular atomic system
implemented in our laboratory and considered in the theoretical
description.
The proposed single-atom model with an arbitrary number of states is
described in Section~\ref{sec:theory}, together with numerical
calculations of the corresponding population dynamics.
Our experimental setup is briefly described in
Section~\ref{sec:experiment}.
The theoretical framework for far-field detection, together with the
experimental and theoretical results for photon counts and
coincidences, is presented in Section~\ref{sec:datavstheory}.
The agreement and discrepancies between experiment and theory are
discussed in Section~\ref{sec:discussion}.
Finally, our conclusions are presented in
Section~\ref{sec:conclusion}.

\section{The atomic system}
\label{sec:system}

\begin{figure}
	\centering
	\includegraphics[scale=0.6]{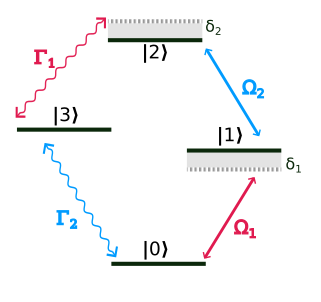}
    \caption{
    Minimal illustration of the diamond configuration for four-wave mixing.
The cyclic process begins with the two-step excitation
$|1\rangle\rightarrow|2\rangle\rightarrow|3\rangle$,
driven by two laser beams with Rabi frequencies $\Omega_{1}$ and
$\Omega_{2}$ and corresponding detunings $\delta_1$ and $\delta_2$,
respectively.
Signal and idler photons are generated through the cascade decay
$|3\rangle\rightarrow|2\rangle\rightarrow|1\rangle$, with decay rates
$\Gamma_1$ and $\Gamma_2$.
The transition colors indicate orthogonal polarizations.
The state $|0\rangle$ represents the lower hyperfine level of the
alkali-atom ground state.}\label{fig:level_scheme}
\end{figure}

Figure~\ref{fig:level_scheme} schematically illustrates the
four-wave mixing process in the so-called diamond configuration
\cite{Willis2008}.
The quantum states $|0\rangle$, $|1\rangle$, $|2\rangle$, and
$|3\rangle$ typically represent hyperfine atomic levels, while the
Zeeman sublevels are neglected.
Although this schematic oversimplifies the dynamics of real atomic
systems, it represents the way FWM has traditionally been modeled:
as a double atomic excitation,
$|0\rangle\rightarrow|1\rangle\rightarrow|2\rangle$,
followed by a cascade decay,
$|2\rangle\rightarrow|3\rangle\rightarrow|0\rangle$.
The two excitation transitions are driven by laser beams with Rabi
frequencies $\Lambda_{01}$ and $\Lambda_{12}$ and 
detunings $\delta_1$ and $\delta_2$, respectively.

Accordingly, traditional single-atom models typically consider up to
five hyperfine atomic levels.

To test the theoretical approach proposed in
Section~\ref{sec:theory}, we present experimental data obtained with
the setup reported in Ref.~\cite{Arias-Tellez:2022} and briefly
described in Section~\ref{setup}.
The setup consists of a correlated photon-pair source based on cold
${}^{87}\mathrm{Rb}$ atomic clouds confined in a magneto-optical trap
(MOT).
For this system, the generic states introduced in
Figure~\ref{fig:level_scheme} correspond to
$|0\rangle\equiv|5S_{1/2},F=1,m_F\rangle$,
$|1\rangle\equiv|5S_{1/2},F=2,m_F\rangle$,
$|2\rangle\equiv|5P_{3/2},F=2,m_F\rangle$,
$|3\rangle\equiv|5P_{3/2},F=3,m_F\rangle$,
$|4\rangle\equiv|5D_{3/2},F=3,m_F\rangle$, and
$|5\rangle\equiv|5P_{1/2},F=2,m_F\rangle$.

The transitions between the $|1\rangle$ and $|3\rangle$
manifolds and between the $|3\rangle$ and $|4\rangle$
manifolds are driven by laser fields at
$780\,\mathrm{nm}$ and $776\,\mathrm{nm}$, respectively,
hereafter referred to as pump~1 and pump~2.
In addition to these transitions, the MOT re-pump laser remains active
during the photon-generation process.
The re-pump laser is right-circularly polarized and drives transitions between
the $|0\rangle$ with $m_F$ and $|2\rangle$ with $m_F+1$ manifolds.
It thereby returns atoms that decay into the lower ground-state
hyperfine manifold
$|0\rangle\equiv|5S_{1/2},F=1,m_F\rangle$
to the excitation cycle.

\begin{figure}
	\centering
	\includegraphics[scale=0.80]{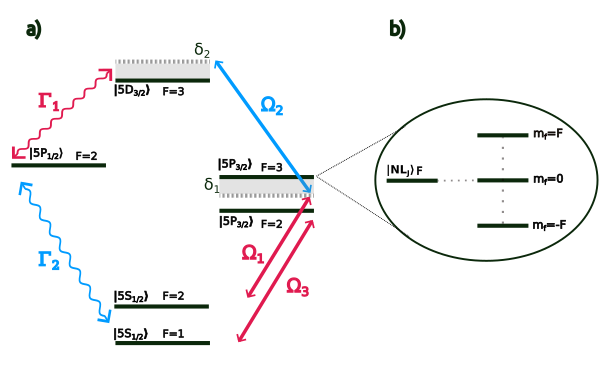}
	\label{fig:32_level_scheme}
    \caption{Illustration of the complete atomic system considered in the model,IOP
corresponding to a diamond configuration of transitions in
${}^{87}\mathrm{Rb}$.
(a) Relevant hyperfine levels, showing the two pump fields that drive
the excitation transitions and the two spontaneous decay transitions,
labeled by their corresponding wavelengths.
(b) All $m_F$ projections included in the model.
For clarity, only a representative subset of the excitation and decay
transitions allowed by the corresponding selection rules is shown.
}
\end{figure}

Consequently, we must consider six hyperfine levels and their corresponding
Zeeman sublevels.
The importance of this is three-fold.
First, all possible excitation and decay transitions allowed by the
selection rules are taken into account.
Second, interference between these transitions is also included.
Third, as explained in Section~\ref{sec:experiment}, the MOT coils
remained energized during the experiment, lifting the degeneracy of
the Zeeman sublevels; this effect is also included in the model.
Accordingly, to track the polarization of the photons throughout the FWM
process, we must consider each $m_F$ projection of the hyperfine states
individually and apply the corresponding selection rules.
As a result of these considerations, the minimum number of states required to model this
nonlinear process within a single-atom description increases to 32,
as shown in Figure~\ref{fig:32_level_scheme}.   

\section{Theory of atomic structure}
\label{sec:theory}

We employ a Hamiltonian that describes the interaction between the
internal atomic states and all relevant laser fields, within the usual
rotating-wave approximation (RWA) and in a reference frame that removes
the explicit time dependence of the driving fields.
The Hamiltonian takes the general form
\begin{equation}
    \mathcal{H}=\mathcal{H}_A+\mathcal{H}_{AF},
\end{equation}
where
\begin{align}
    \mathcal{H}_A &=
    \sum_{i}^{n_l} e_{ii}\sigma_{i,i},\\
    \mathcal{H}_{AF} &=
    \sum_{\{i,j\}}\Omega_{i,j}
    \left(\sigma_{i,j}+\sigma_{j,i}\right),
\end{align}
with $\sigma_{i,j}=\ket{i}\bra{j}$ denoting the standard atomic
operators that couple state $\ket{j}$ to state $\ket{i}$, and $e_{ii}$ is the energy of level $i$.
The indices $i$ and $j$ label the atomic quantum states, which are of
the form $\left|I,J,F,m_F\right\rangle$, where $I$ is the nuclear
angular momentum, $J$ is the electronic angular momentum, $F$ is the
total angular momentum, and $m_F$ is its projection, with
$-F\leq m_F\leq F$.
In the reference frame used to eliminate the explicit time dependence
of the Hamiltonian, the diagonal energies $e_{ii}$ are linear combinations of the laser detunings.
The curly brackets under the summation sign indicate that $i>j$, and
the indices run over all pairs of atomic levels coupled by an external
laser field, with Rabi frequency
\begin{equation}
\Omega_{i,j} = \frac{p_{i,j}}{\hbar}
\sqrt{\frac{2\mu_0 c P_{i,j}}{\pi r^2}},
\label{eq:rabi00}
\end{equation}
where $r$ is the laser spot radius, $p_{i,j}$ is the transition dipole
matrix element, $c$ is the speed of light in vacuum, $\mu_0$ is the
vacuum permeability, and $P_{i,j}$ is the input power of the laser
field coupling the corresponding transition. To simplify the notation, we denote the Rabi frequencies of pump~1
and pump~2 by $\Omega_1=\Omega_{1,3}$ and
$\Omega_2=\Omega_{3,4}$, respectively, and their corresponding
powers by $P_1=P_{1,3}$ and $P_2=P_{3,4}$.
Additionally, we denote the Rabi frequency of the re-pump laser by
$\Omega_3=\Omega_{0,2}$ and its power by $P_3=P_{0,2}$.
The model accounts for both the polarization of the laser fields and the
corresponding selection rules for the atomic dipole transitions.

Spontaneous emission is also included, and the density matrix of the
system is therefore obtained from the Lindblad master equation
\begin{equation}
    \dot{\rho}
    = \frac{1}{i\hbar}[\mathcal{H},\rho]
    + \mathcal{L}[\rho],
    \label{eq:GeneralMasterEq}
\end{equation}
where
\begin{equation}
    \mathcal{L}[\rho]
    = \sum_{\{i,j,k,l\}}^{n_l} \gamma_{i,j,k,l}
    \left(
    \sigma_{k,l}\rho\sigma_{i,j}^\dagger
    -\frac{1}{2}
    \left\{\sigma_{i,j}^\dagger\sigma_{k,l},\rho\right\}
    \right).
\end{equation}
The sum runs over all pairs of transitions
$\{i,j\}$ and $\{k,l\}$ for which spontaneous decay is allowed,
with $\gamma_{i,j,k,l}$ denoting the corresponding dissipative
coupling between the two transitions.
The coefficients $\gamma_{i,j,k,l}$ are given by
\begin{equation}
\gamma_{i,j,k,l}
=
\frac{1}{6\pi\epsilon_0\hbar}
\kappa_{i,j,k,l}^3
\boldsymbol{p}_{i,j}\cdot\boldsymbol{p}_{k,l}^*,
\end{equation}
where
$\kappa_{i,j,k,l}=(\Delta k_{i,j}+\Delta k_{k,l})/2$
is the average of the wave numbers $\Delta k_{i,j}$ and
$\Delta k_{k,l}$ associated with the $\{i,j\}$ and $\{k,l\}$
transitions, respectively.

The dipole matrix element $\boldsymbol{p}_{i,j}$ for the
$\{i,j\}$ transition is a vector calculated using the Wigner-Eckart theorem,
where the indices $i$ and $j$ label the quantum states
\(
|i\rangle\equiv\left|I,J,F,m_F\right\rangle
\),
\(
|j\rangle\equiv
\left|I,J^\prime,F^\prime,m_F^\prime\right\rangle,
\)
respectively. Thus,
the dipole matrix element
is calculated using the Wigner-Eckart theorem as
\begin{align}
\boldsymbol{p}_{i,j}
=&-e
\left\langle I,J,F,m_F\right|
\boldsymbol{r}
\left|I,J^\prime,F^\prime,m_F^\prime\right\rangle
\nonumber\\
=&\frac{p_{ij}}{\sqrt{2J+1}}
\sum_{m_J=-J}^{J}
\sum_{m_{J^\prime}=-J^\prime}^{J^\prime}
\sum_{m_I=-I}^{I}
C_{J^\prime,m_{J^\prime},I,m_I}^{F^\prime,m_F^\prime *}
C_{J,m_J,I,m_I}^{F,m_F}
\nonumber\\
&\times
\bigg(
C_{J,m_J,1,-1}^{J^\prime,m_{J^\prime}}\boldsymbol{e}_L
-
C_{J,m_J,1,1}^{J^\prime,m_{J^\prime}}\boldsymbol{e}_R
+
C_{J,m_J,1,0}^{J^\prime,m_{J^\prime}}\boldsymbol{e}_k
\bigg).
\end{align}
Here, $C_{J,m_J,I,m_I}^{F,m_F}$ denotes a Clebsch--Gordan
coefficient,
$\boldsymbol{e}_L=(\boldsymbol{e}_H+i\boldsymbol{e}_V)/\sqrt{2}$ and
$\boldsymbol{e}_R=(\boldsymbol{e}_H-i\boldsymbol{e}_V)/\sqrt{2}$
are the left- and right-circular polarization unit vectors,
respectively, while $\boldsymbol{e}_H$, $\boldsymbol{e}_V$, and
$\boldsymbol{e}_k$ are unit vectors along the horizontal, vertical,
and propagation directions, respectively.
The reduced dipole matrix element is incorporated into this expression
through $p_{ij}$.
For simplicity, in Fig.~\ref{fig:32_level_scheme}, we use
$\Gamma_1$ and $\Gamma_2$ as collective labels for the sets of
dissipative couplings associated with the corresponding hyperfine
manifolds.
Specifically, $\Gamma_1$ represents all allowed dissipative couplings
between the Zeeman sublevels of
$\ket{5D_{3/2},F=3}$ and $\ket{5P_{1/2},F=2}$,
whereas $\Gamma_2$ represents those between the Zeeman sublevels of
$\ket{5P_{1/2},F=2}$ and $\ket{5S_{1/2},F=1,2}$.

With a total of 32 atomic levels, deriving explicit analytical
expressions becomes cumbersome and inefficient, and the resulting
equations of motion are extremely involved.
For this reason, atomic models developed to address this type of
problem typically consider the smallest possible number of levels.
However, to obtain the equations of motion and their solutions without
resorting to this simplification, we employ the MulAtoLEG
\textit{Mathematica} package~\cite{Yanes-Thomas:2026}.
The package requires only the specification of the relevant atomic
levels, external coupling fields, and physical parameters to
automatically generate the Lindbladian and the corresponding
differential equations.

\subsection{Numerical solution of equations}

In order to solve Eq.~\eqref{eq:GeneralMasterEq}, we employed the method detailed in \cite{Yanes-Thomas:2025}, in which the master equation is projected onto a suitable matrix basis, yielding a linear system of equations for the expansion coefficients. Once this system is solved, the expectation value of any arbitrary atomic operator can be calculated as a function of time. To compare with the experimental results, we use the steady-state expectation values.

It is estimated that the MOT coils produce magnetic fields up to $1\,\mathrm{G}$ within the region where the atoms interact with the pump beams, Section~\ref{sec:experiment}.
To assess the effect of the resulting spatial variation of the
Hamiltonian due to the Zeeman shifts experienced by atoms across the cold cloud, we solved the equations multiple times, assigning a random magnetic-field magnitude within this range and averaging the resulting observables. Between $0$ and $0.35\,\mathrm{G}$ we found the effect of these magnetic-field variations to be minimal and to be consistent with experimental results. For larger magnetic fields, however, the theoretical results become strongly dependent on the field strength. We concluded that the atoms generating the bulk of the light observed in our experiments are located within a radius of 0.1 mm from the center of the pump beams. Thus, the theoretical results presented in Section~\ref{sec:datavstheory} neglect the local magnetic-field
variation within the atomic cloud. It would nevertheless be interesting to perform similar calculations and compare them with experimental measurements at local magnetic fields above $0.35\,\mathrm{G}$, where the increased Zeeman splitting
is expected to have a more significant effect on the photon-pair correlations.

The simulations were performed over parameter ranges chosen to match
those explored experimentally.
The reduced dipole matrix elements for the transitions
$\left|5S_{1/2}\right\rangle\rightarrow\left|5P_{3/2}\right\rangle$
($780\,\mathrm{nm}$),
$\left|5P_{3/2}\right\rangle\rightarrow\left|5D_{3/2}\right\rangle$
($776\,\mathrm{nm}$),
$\left|5D_{3/2}\right\rangle\rightarrow\left|5P_{1/2}\right\rangle$
($762\,\mathrm{nm}$), and
$\left|5P_{1/2}\right\rangle\rightarrow\left|5S_{1/2}\right\rangle$
($795\,\mathrm{nm}$)
of ${}^{87}\mathrm{Rb}$ were set to
$5.956\,e a_B$, $0.787\,e a_B$, $1.616\,e a_B$, and
$4.11\,e a_B$, respectively, where $e$ is the elementary charge and
$a_B$ is the Bohr radius \cite{Safronova2004,Yanes-Thomas:2025}.

We recall that pump~1 and pump~2 drive the
$\ket{5S_{1/2},F=2}\rightarrow\ket{5P_{3/2},F=3}$
and
$\ket{5P_{3/2},F=3}\rightarrow\ket{5D_{3/2},F=3}$
transitions, respectively, with corresponding input powers
$P_1$ and $P_2$.
Both beams were experimentally set to a FWHM of
$1.1\,\mathrm{mm}$, as described in
Section~\ref{sec:experiment}.
However, spot radii of $1.0\,\mathrm{mm}$ and $1.5\,\mathrm{mm}$ for pump 1 and pump 2, respectively, yielded the best
agreement between experimental and theoretical results.
In both cases, their powers were varied up to
$P_1 = 1.5\,\mathrm{mW}$
and $P_2 = 18\,\mathrm{mW}$.
According to
Eq.~\eqref{eq:rabi00}, these values correspond
to Rabi frequencies of $\Omega_1 = 13\,\mathrm{MHz}$
and $\Omega_2 = 22\,\mathrm{MHz}$ respectively. 
The detuning of the \(780\,\mathrm{nm}\) laser was kept constant at
\(\delta_1=-70\,\mathrm{MHz}\), and the detuning of the
\(776\,\mathrm{nm}\) laser was set to \(\delta_2=71.5\,\mathrm{MHz}\)
so that the two-photon detuning $\Delta=\delta_1+\delta_2$
remained fixed at
\(\Delta=1.5\,\mathrm{MHz}\).
The re-pump laser was kept at  $P_3=10\,\mathrm{mW}$ with
right-circular polarization and tuned to resonance with the transition.

\subsection{Population dynamics}
Cooling and trapping alkali atoms in a MOT follows a well-established protocol in which a ``re-pump''  laser ensures that the atoms are subject to the laser force by depopulating their lowest ground level and returning them to the excitation cycle. In the case of ${}^{87}\mathrm{Rb}$, the re-pump laser is resonant with the $\left|5S_{1/2},F=2\right\rangle \rightarrow
\left|5P_{3/2},F=3\right\rangle$ transition. In the present section we study the influence of this process on the populations of the 32 atomic states involved within the diamond scheme analyzed in this article.

\begin{figure}[h]
\includegraphics[width=0.245\textwidth]{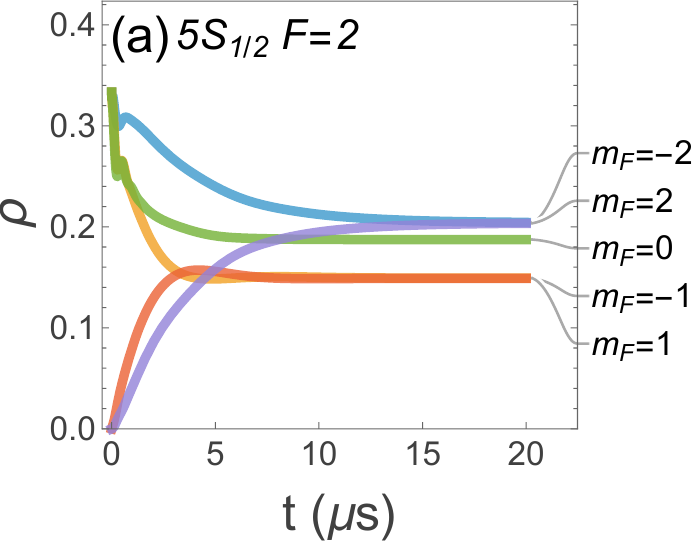}
\includegraphics[width=0.245\textwidth]{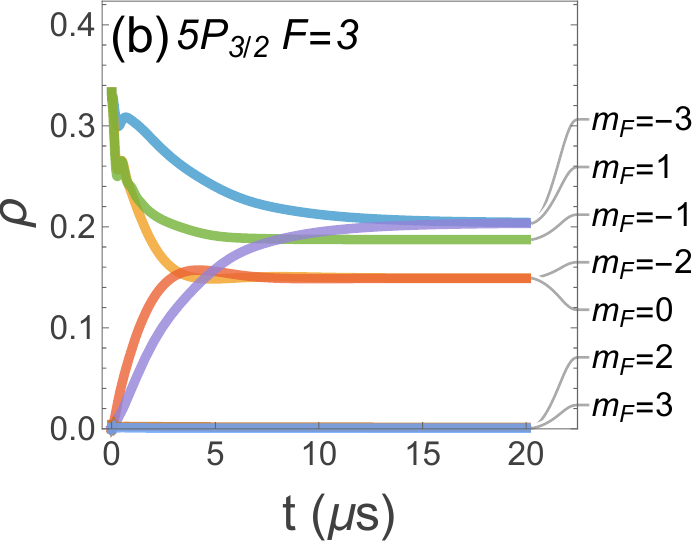}
\includegraphics[width=0.245\textwidth]{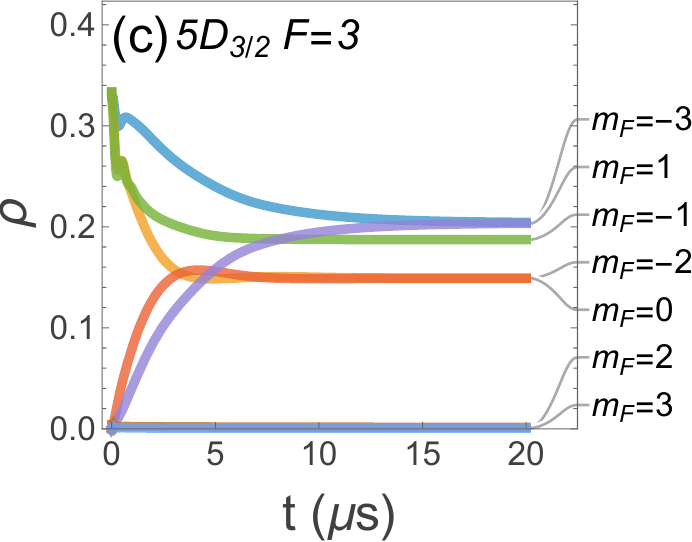}
\includegraphics[width=0.245\textwidth]{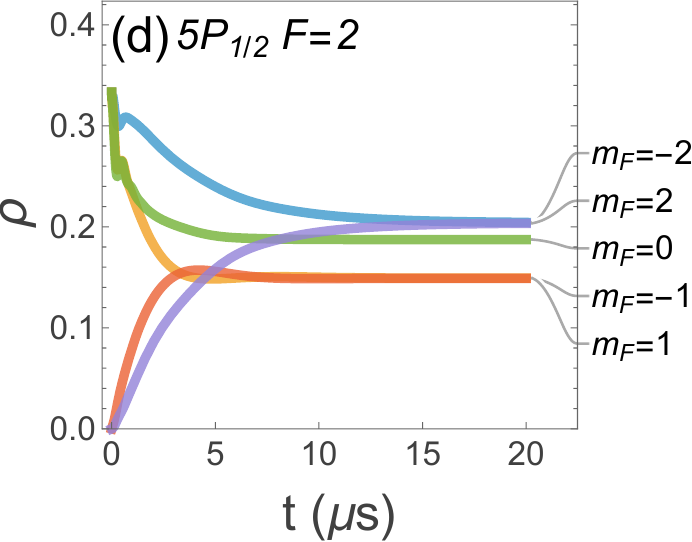}\\
\includegraphics[width=0.245\textwidth]{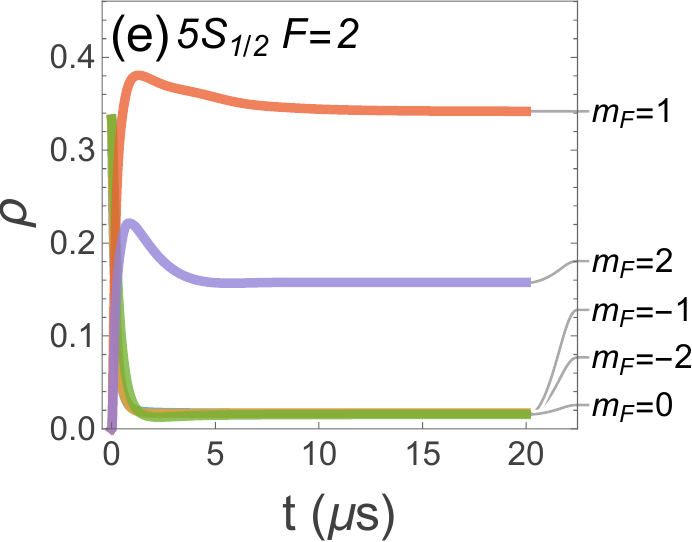}
\includegraphics[width=0.245\textwidth]{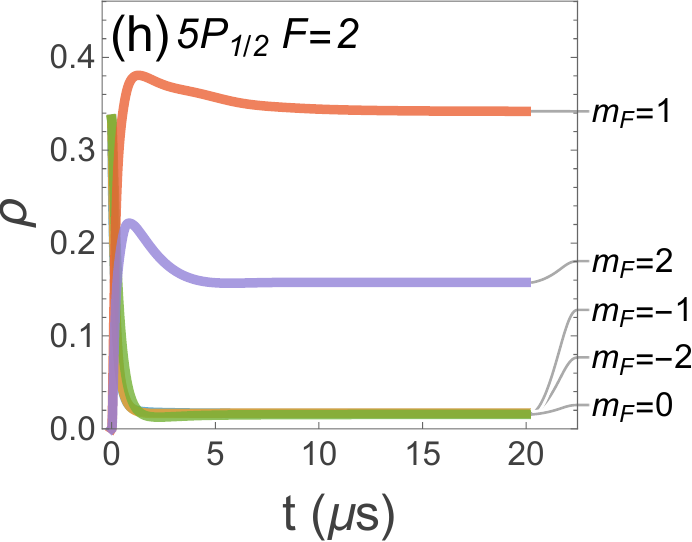}
\includegraphics[width=0.245\textwidth]{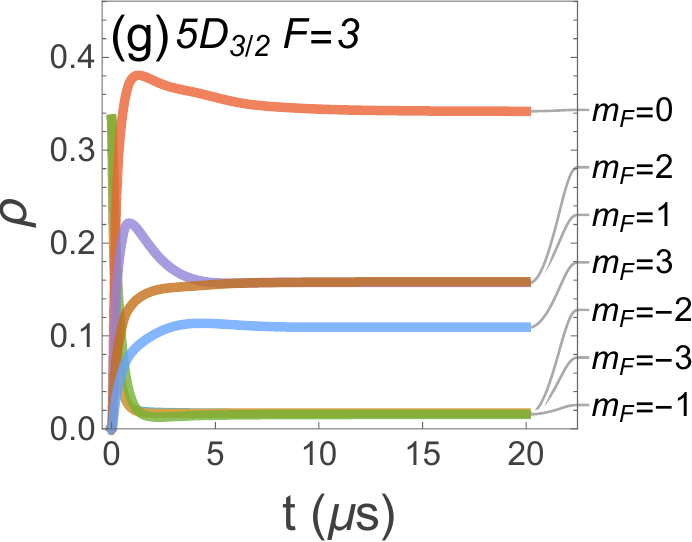}
\includegraphics[width=0.245\textwidth]{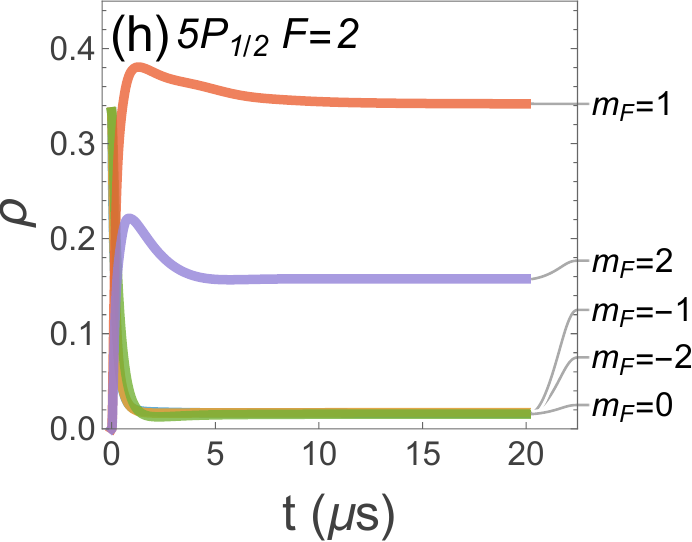}\\
\includegraphics[width=0.246\textwidth]{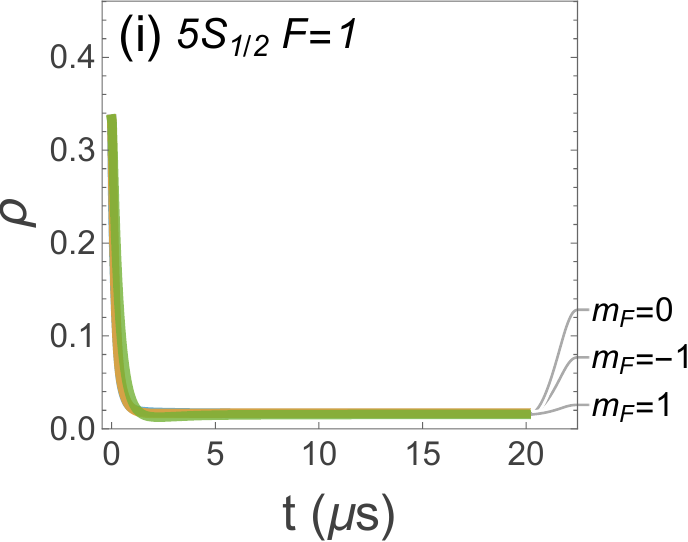}
\includegraphics[width=0.246\textwidth]{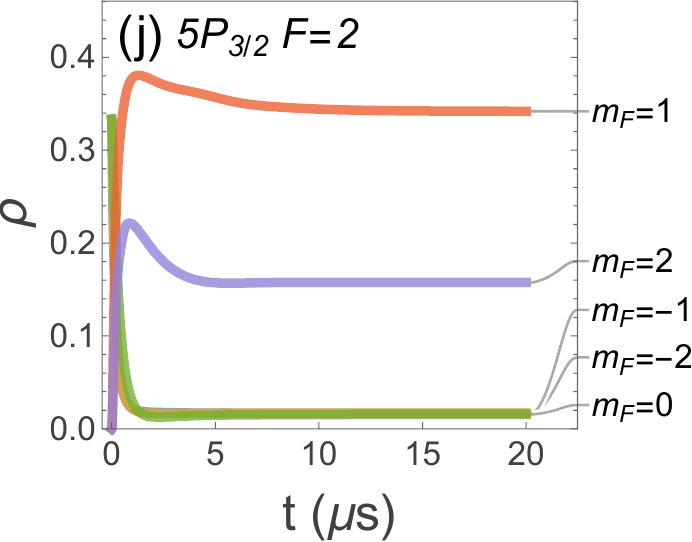}
\caption{Populations of all possible $m_F$ projections as a function of time.
Panels (a)--(d) show the populations of the $m_F$ projections
of the $5S_{1/2}\,F=2$, $5P_{3/2}\,F=3$,
$5D_{3/2}\,F=3$, and $5P_{1/2}\,F=2$ levels
without the re-pump laser.
Panels (e)-(h) show the corresponding populations
when the re-pump laser is included.
Panels (i) and (j) show the populations of the auxiliary levels
$5S_{1/2}\,F=1$ and $5P_{3/2}\,F=2$,
which participate in the re-pump process. Results obtained via simulation with the MulAtoLEG
\textit{Mathematica} package. }
\label{fig:repump24-32}
\end{figure}

To begin, we explore the importance of accounting for all the $m_F$ projections. We then introduce the re-pump laser to the full, one-atom model. Panels (a)--(d) of Figure~\ref{fig:repump24-32} display the dynamics of the quantum levels
(a) $\left|5S_{1/2},F=2\right\rangle$,
(b) $\left|5P_{3/2},F=3\right\rangle$,
(c) $\left|5D_{3/2},F=3\right\rangle$, and
(d) $\left|5P_{1/2},F=2\right\rangle$
for all corresponding values of
$m_F=-F,$ $-F+1, $ $ \ldots , $ $F-1,$ $F$,
as a function of time without the re-pump laser.
For these curves, the pump beams were assigned horizontal (H) and vertical (V) linear polarizations with $P_1=0.5\,\mathrm{mW}$ and
$P_2=15\,\mathrm{mW}$, respectively.   
Figures~\ref{fig:repump24-32} (e)--(h) show the populations of the corresponding
levels when the re-pump scheme is included while keeping the parameters of the FWM pump beams unchanged. In this case, the auxiliary levels
$\left|5S_{1/2},F=1\right\rangle$ and
$\left|5P_{3/2},F=2\right\rangle$
[shown in panels (i) and (j)]
must also be considered, as they are coupled by a
right-circularly polarized re-pump laser with fixed power
$P_3=10\,\mathrm{mW}$.
As expected, this coupling depopulates the
$\left|5S_{1/2},F=1\right\rangle$ level, as shown in panel (i).
The populations of the magnetic sublevels of
$\left|5P_{3/2},F=2\right\rangle$, however, do not exhibit the
distribution expected from optical pumping by a circularly polarized
re-pump laser alone.
Instead, they are strongly influenced by the dynamics of the
surrounding atomic levels and the competing optical transitions.

Comparison between the first and second rows of plots in Figure \ref{fig:repump24-32} reveals that the relevant magnetic sublevels change dramatically when the re-pump
laser is included.
Without the re-pump laser, the lowest-$m_F$ projections are populated slightly more than the remaining projections and all populations saturate at a comparable magnitude.
When the re-pump laser is included, however, the
$m_F=0$ and $m_F=1$ projections become dominant, giving rise to a clear
hierarchy among the magnetic sublevels.
Except for the ground state
$\ket{5S_{1/2},F=2}$, all atomic levels exhibit at least two, and in some cases
more, significantly populated $m_F$ projections.

Since the polarization of the emitted photons is determined by the
specific magnetic sublevels involved in each transition, a model that
considers only four atomic levels is insufficient for a truly accurate description of
the polarization-resolved emission. Accordingly, our model explicitly includes all magnetic sublevels
as well as the re-pump scheme.

\section{Experiment} 
\label{sec:experiment}
\begin{figure}[h]
\includegraphics[width=0.8\textwidth]{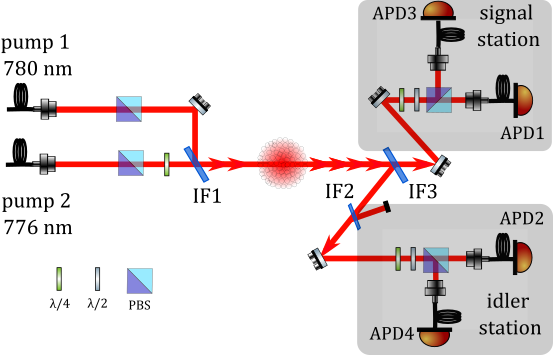}
\caption{Schematic of the experimental apparatus.
Pump~1 and pump~2 (left) are delivered to the main experiment through
optical fibers and stabilized using two saturated-absorption
spectroscopy setups.
Their polarizations are prepared to be linear and orthogonal, and the
beams are overlapped at interference filter IF1 so that they
co-propagate through the center of the magneto-optical trap.
A combined beam containing four wavelengths exits the vacuum chamber
on the opposite side.
Interference filter IF3 transmits only the signal photons, which are
directed to the first quantum-tomography station equipped with APD1
and APD3.
The remaining pump light is subsequently filtered out by IF2, which
transmits only the idler photons and directs them to the second
quantum-tomography station equipped with APD2 and APD4.  }
\label{setup}
\end{figure}

A simplified schematic of our experimental setup is shown in
Figure~\ref{setup}. It is a time-correlated photon-pair source based on cold atomic samples prepared in a magneto-optical trap (MOT). For these data, clouds typically had an approximate diameter of 5 mm and  $\approx 5\times10^7$ atoms at a temperature of $1.5\,\mathrm{mK}$~\cite{Arias-Tellez:2022}. The photon-collection array is equipped with two polarization-analysis stations for quantum-state tomography.   

The pump beams are provided by two extended-cavity diode lasers
stabilized through saturated-absorption spectroscopy (setups not shown). For the data presented in Section~\ref{sec:datavstheory}, the detunings are $\delta_1=-70\,\mathrm{MHz}$ and
$\Delta=+2\,\mathrm{MHz}$.
Both beams are collimated to an approximate FWHM of $1.1\,\mathrm{mm}$.
Pump powers are varied over the ranges
$P_1=0.1$--$1\,\mathrm{mW}$ and
$P_2=1$--$16\,\mathrm{mW}$. Their polarizations are set to $H$ and $V$, respectively. The two beams are combined using interference filter IF1 (Semrock LL01-780-25) and co-propagated toward the center of the atomic cloud. 

Photons are generated using a pulsed sequence with a duty cycle
that maximizes signal and idler photon counts.
During each cycle, the MOT is loaded for $500\,\mu\mathrm{s}$ and, during the following $200\,\mu\mathrm{s}$, the cooling beams are switched off whilst the pump beams are turned on. The MOT coils are left on throughout both stages of the cycle, providing a magnetic gradient of 20 $\mathrm{G\cdot cm}^{-1}$.

\begin{figure}[!htbp]
\begin{minipage}{0.4\textwidth}
\includegraphics[width=0.98\textwidth]{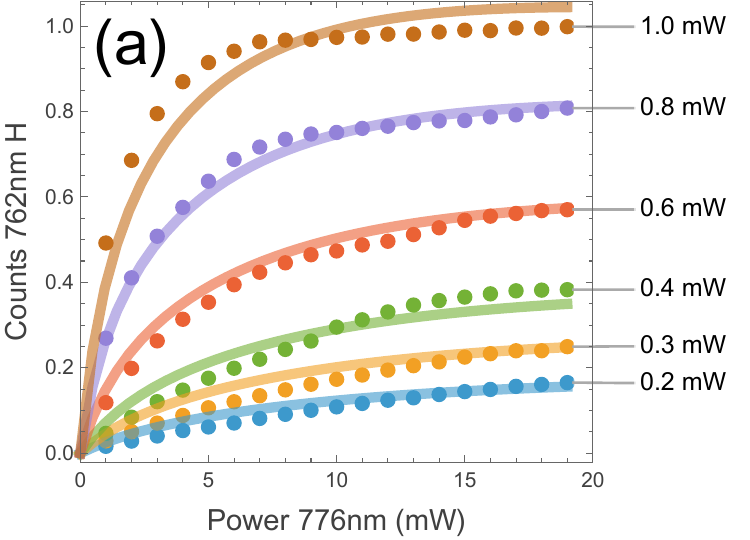}
\includegraphics[width=0.98\textwidth]{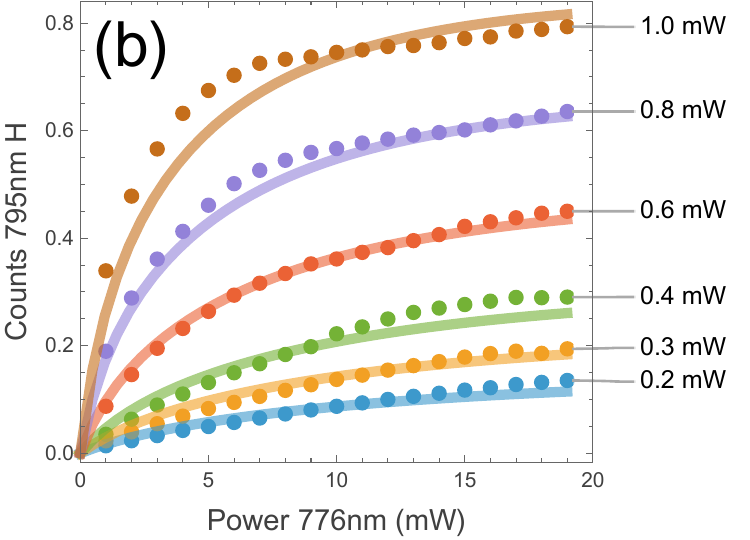}
\includegraphics[width=0.98\textwidth]{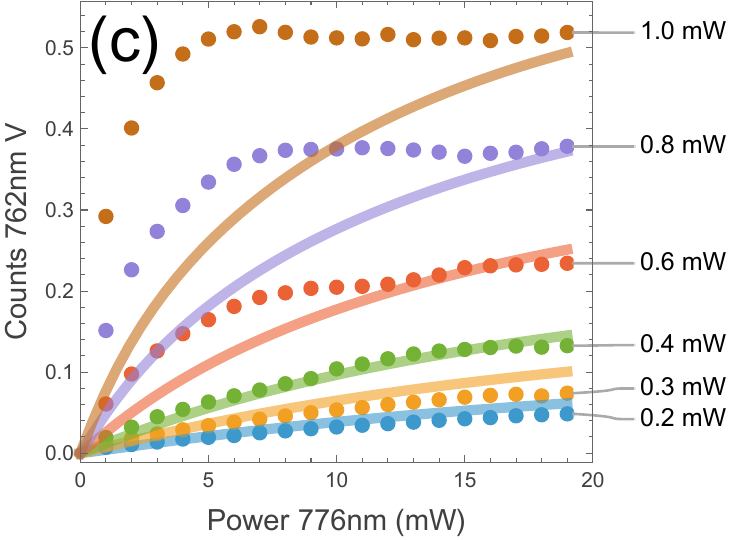}
\includegraphics[width=0.98\textwidth]{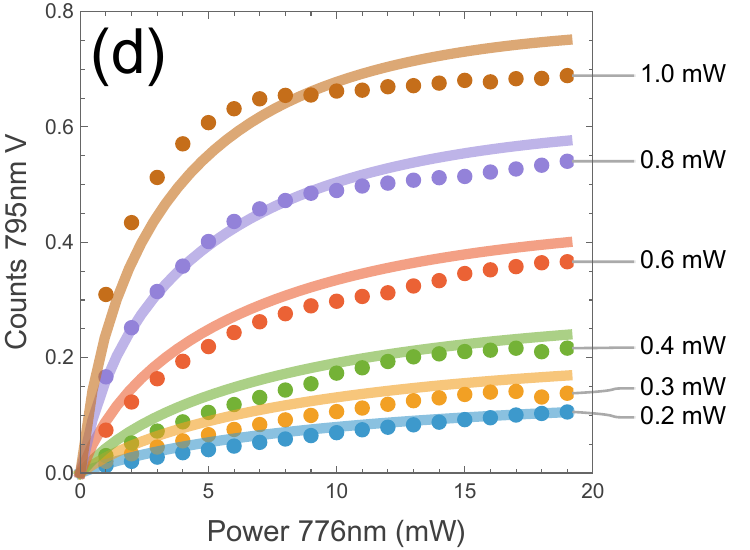}
\end{minipage}
\begin{minipage}{0.4\textwidth}
\includegraphics[width=0.98\textwidth]{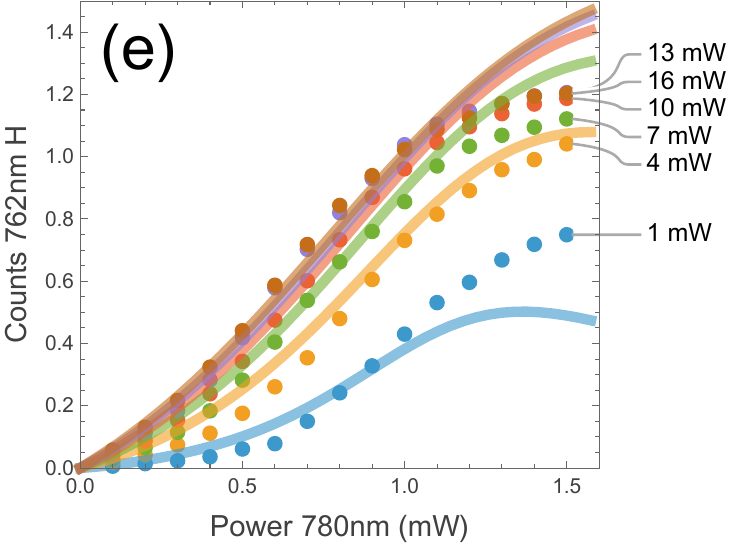}
\includegraphics[width=0.98\textwidth]{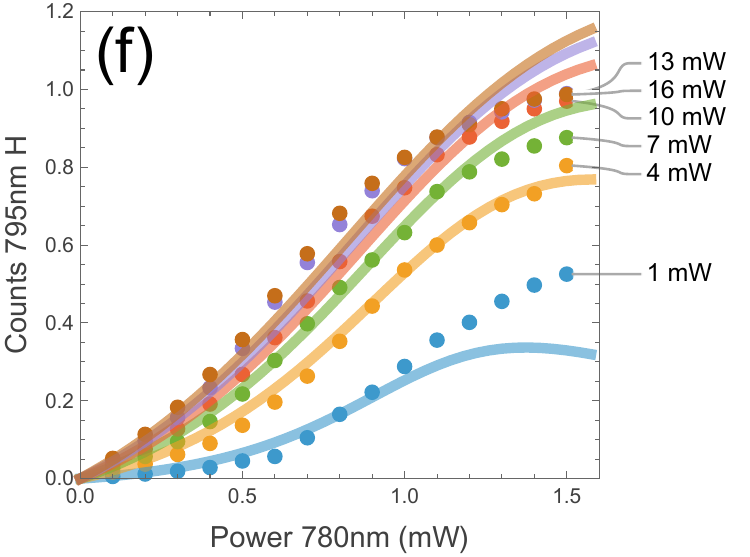}
\includegraphics[width=0.98\textwidth]{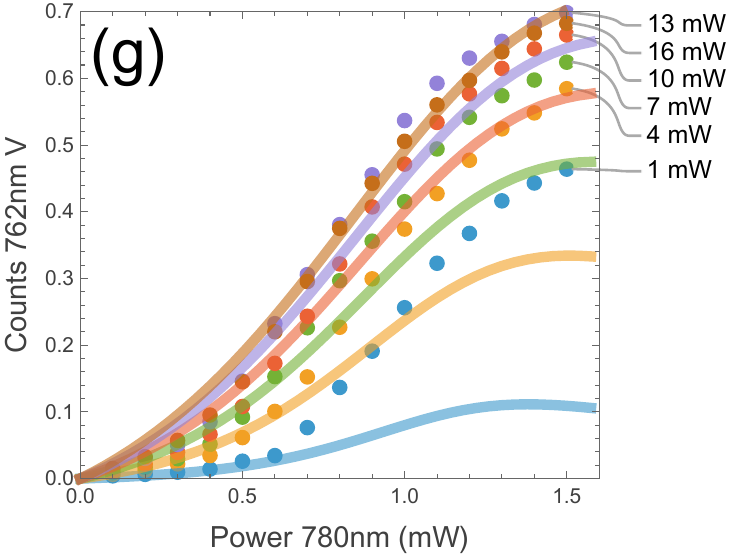}
\includegraphics[width=0.98\textwidth]{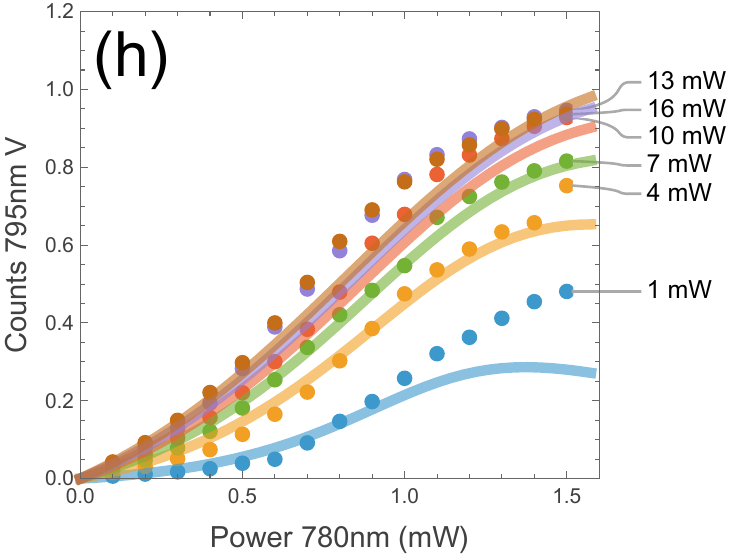}
\end{minipage}

\caption{Experimental and theoretical results for the normalized photon counts to the highest value in (a),
organized by polarization and wavelength,
as functions of $P_1$ ($780\,\mathrm{nm}$) and $P_2$ ($776\,\mathrm{nm}$).
The theoretical calculations are based on the 32-level model.
Panels (a), (b), (e), and (f) show the normalized counts for
horizontal (H) polarization,
while panels (c), (d), (g), and (h) show the normalized
counts for vertical polarization.
When the normalized counts are shown as a function of $P_1$,
multiple curves are plotted for different values of $P_2$, and vice versa.
}
\label{fig:countsP1}
\end{figure}

\begin{figure}[h!htbp]
\begin{minipage}{0.48\textwidth}
\includegraphics[width=0.98\textwidth]{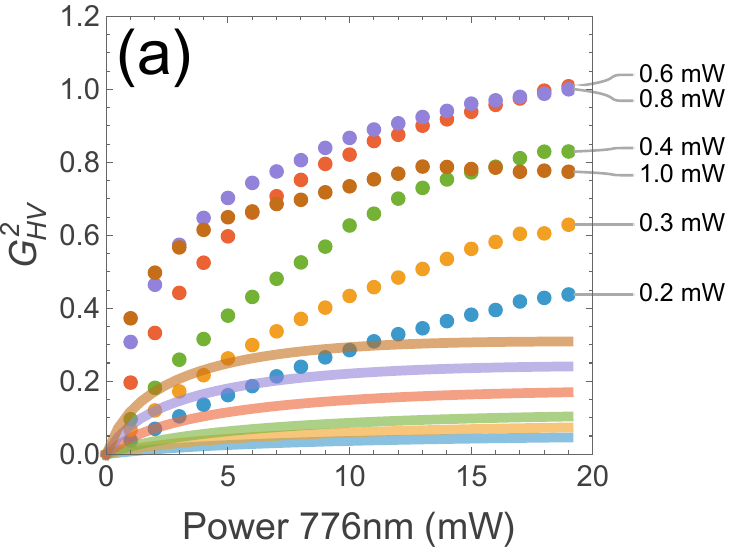}
\includegraphics[width=0.98\textwidth]{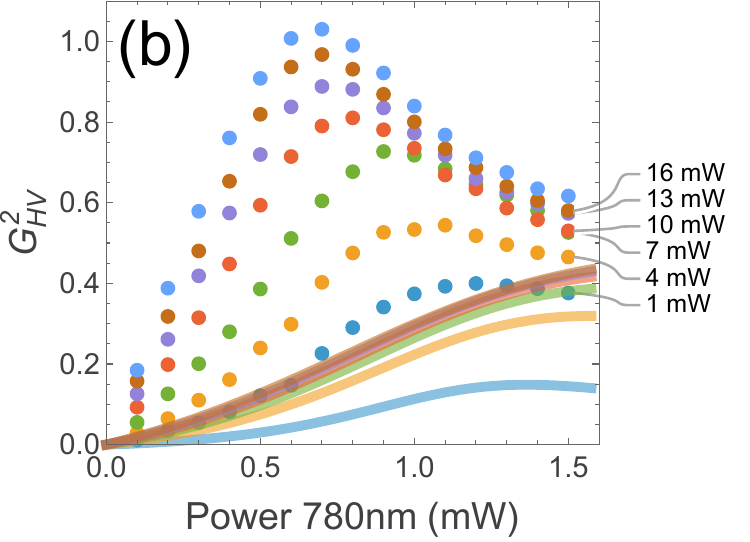}
\end{minipage}
\begin{minipage}{0.48\textwidth}
\includegraphics[width=0.98\textwidth]{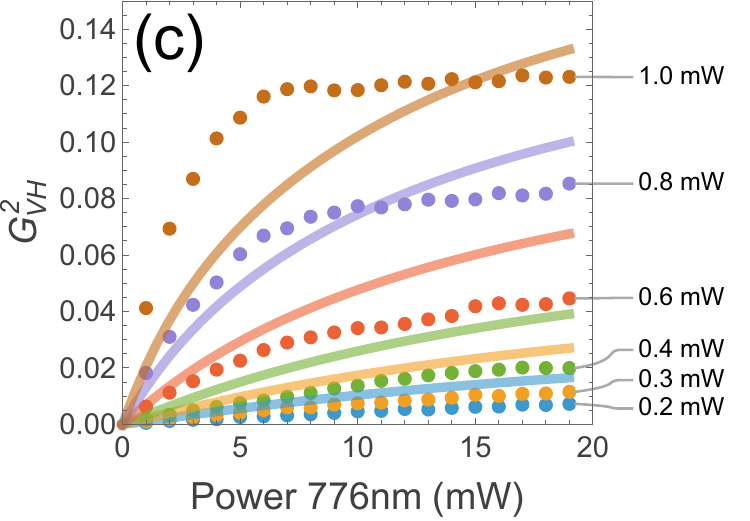}
\includegraphics[width=0.98\textwidth]{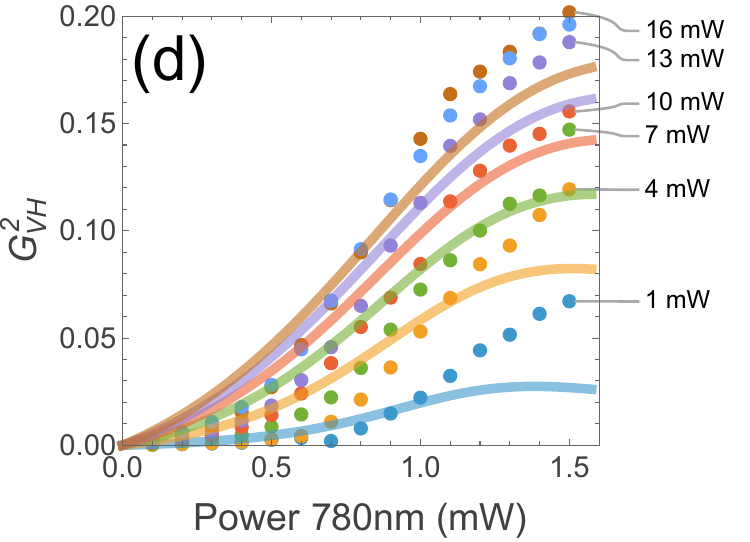}
\end{minipage}
\caption{
Normalized photon coincidences organized by polarization and wavelength
as functions of $P_1$ ($780\,\mathrm{nm}$) and
$P_2$ ($776\,\mathrm{nm}$).
Panels (a) and (b) show the correlation $G_{HV}^{(2)}$ between
horizontally (H) polarized photons emitted
in the $762\,\mathrm{nm}$ transition and
vertically (V) polarized photons emitted
in the $795\,\mathrm{nm}$ transition.
In contrast, panels (c) and (d) show
the correlation $G_{VH}^{(2)}$ between
vertically (V) polarized photons emitted
in the $762\,\mathrm{nm}$ transition and
horizontally (H) polarized photons emitted
in the $795\,\mathrm{nm}$ transition.
When the photon coincidences are shown as a function of $P_1$,
multiple curves are plotted for different values of $P_2$, and vice versa.
}
\label{fig:coincidencesP1}
\end{figure}

Since the two pump beams propagate collinearly, the signal
($762\,\mathrm{nm}$) and idler ($795\,\mathrm{nm}$) photons are
emitted along the same direction as the pump beams, in accordance
with momentum conservation in the FWM process.
To separate the signal and idler photon fluxes from the pump fields,
two pairs of interference filters, IF2 (Semrock LL01-780-25) and
IF3 (Semrock LL01-808-25), are employed.
After the pump light is filtered out, the generated photons are
directed to their respective detection stations.

Each station is equipped with a polarization analyzer consisting of a
$\lambda/4$ wave-plate, a $\lambda/2$ wave-plate, and a polarizing
beam-splitter cube (PBS), followed by two avalanche photodiodes
(APDs, ID Quantique ID120-500-800) coupled through single-mode optical
fibres. The wave-plates at each station are configured such that
$H$-polarized light is transmitted by the PBS, whereas $V$-polarized
light is reflected. The counts detected by the four avalanche photodiodes are processed through AND-logic circuits and registered by a time-to-digital converter (ID Quantique ID800-TDC).

\section{Far field measurements and experimental data}
\label{sec:datavstheory}

Counts and coincidences are obtained from photodetection measurements of the emitted electromagnetic field.
To relate the expectation values of the atomic operators $\sigma_{ij}$ to these measurements, we employ the far-field
operator~\cite{Lehmberg:1970}
\begin{equation}
\boldsymbol{E}_{\text{s},ij}^{(+)}(\boldsymbol{R},t)=
\frac{\hbar}{\epsilon_0c^2}
\frac{\boldsymbol{p}_{ij}
    -\hat{\boldsymbol{R}}(\hat{\boldsymbol{R}}
    \cdot \boldsymbol{p}_{ij})}{R}
\Delta_{ij}^2
\exp\left[i\left(R_{0}-R\right)\kappa_{ij}\right]
\sigma_{ij}(t)\, .
\label{eq:farfield}
\end{equation}
Here, $\boldsymbol{R}$ denotes the detector position,
$\boldsymbol{R}_0=\boldsymbol{r}_0-\boldsymbol{R}$,
$\boldsymbol{r}_0$ is the position of the atom,
$\boldsymbol{p}_{ij}$ is the electric-dipole moment associated with the
transition between levels $i$ and $j$, and
$\Delta_{ij}=c\kappa_{ij}$ is the transition frequency,
where $\kappa_{ij}$ is the corresponding wave number.
The transition is mediated by the operator $\sigma_{ij}(t)$.
The subscript $s$ indicates that this is the electric field
scattered by the atom, as opposed to the driving laser fields,
which are filtered out in the experiment. Equation~\eqref{eq:farfield} is used to calculate both \textbf{photon counts} and \textbf{coincidences}
for direct comparison with the experimental data.

\subsection{Counts}
\label{sub:counts}

Photon counts are modeled in terms of the electric-field intensity at the detector position, which is determined by the first-order correlation function of the corresponding field operators. Throughout this work, the detection time is taken to be $t=0$.

The corresponding quantity is the first-order correlation function,
\begin{equation}
\label{eq:g1}
G^{(1)}_{s,ij,\epsilon}(R)
=
\expval{E^\dagger_{s,ij,\epsilon}(R)
E_{s,ij,\epsilon}(R)
},
\end{equation}
where $E_{s,ij,\epsilon}(R)=E_{s,ij,\epsilon}(R)\cdot e_\epsilon$ denotes the far-field electric-field
operator associated with the $i\rightarrow j$ transition, projected
onto the polarization direction $\epsilon\in\{H,V\}$.

Figure~\ref{fig:countsP1} compares the experimental and theoretical
results for the photon counts recorded by the four avalanche
photodiodes shown in Figure~\ref{setup}.
The colored dots represent the experimental data; their uncertainty
bars are smaller than the symbols and are therefore not visible.
The colored curves correspond to the theoretical results calculated
using Equation~\eqref{eq:g1}.
Recall that, at each detection station, the transmission port of the
PBS directs horizontally polarized light to one avalanche photodiode,
whereas the reflection port directs vertically polarized light to the
other.
All data have been normalized with respect to the maximum number of
counts recorded in Figure~\ref{fig:countsP1}(a).

Panels (a)--(d) show the photon counts as a function of increasing
$P_2$, with each curve corresponding to a fixed value of $P_1$.
Conversely, panels (e)--(h) show the photon counts as a function of
increasing $P_1$, with each curve corresponding to a fixed value of
$P_2$. 

Panels (a), (c), (e), and (g) of Figure~\ref{fig:countsP1} display the results obtained from the signal station. Panels (a) and (c) show the photon counts at the signal wavelength ($762\,\mathrm{nm}$) as a function of $P_2$, whereas panels (e) and (g) show the corresponding results as a function of $P_1$. The approach to saturation is more pronounced as a function of $P_2$ than as a function of $P_1$. Theory and experiment show excellent agreement in panel (a). For panel (c), the agreement deteriorates as $P_1$ increases. The subsaturation value of the counts for H-polarized photons in panel (a) is approximately twice that for V-polarized photons in panel (c). As a function of $P_1$, the saturation value is approached more gradually, as can be seen in panels (e) and (g). Interestingly, the agreement between theory and experiment in these panels behaves differently. The agreement is better for higher values of $P_2$, with a clear compression of the experimental curves near their saturation values. This could be caused by a systematic error associated with the polarization quality of the light transmitted by the PBS\footnote{Thorlabs part no. PBS102; 95 \% efficiency for reflection (H), 90 \% efficiency for transmission (V). }. 

Counts detected at the idler station are presented in
Fig.~\ref{fig:countsP1}(b), (d), (f), and (h), corresponding to the
detection of horizontally (H), vertically (V), horizontally (H), and
vertically (V) polarized photons at $795\,\mathrm{nm}$, respectively.
Panels (b) and (d) show good agreement between experiment and theory
over the entire range of $P_1$ powers investigated. Although a
difference in the saturation values between theory and experiment is
still observed, it is less pronounced than in the corresponding signal
counts. Likewise, panels (f) and (h), which show the dependence of the
idler counts on $P_1$, exhibit behavior similar to that observed at
the signal station, except that the saturation values are
significantly larger and much closer to one another. 

The data presented above were acquired under conditions that maximize the generated photons. Consequently, the
measured counts include not only photons generated through four-wave
mixing but also photons originating from competing processes, such as
amplified spontaneous emission~\cite{Malcuit:1985}. At least within the
range of experimental parameters explored here, a polarization analysis
of the detected counts alone is therefore insufficient to distinguish
between these contributions.

\subsection{Coincidences}
\label{sub:coincidences}

A coincidence event corresponds to the detection of two photons
within a short window of time. The probability of such an event is proportional to the conditional
probability of detecting the photon associated with
$E_{s,kl,\epsilon^\prime}(R_2)$, given that the photon associated with
$E_{s,ij,\epsilon}(R_1)$ has already been detected.
This probability is described by the second-order correlation function,
\begin{equation}
G^{(2)}_{s,ij,\epsilon ;kl, \epsilon^\prime}(R_1,R_2)
=\expval{
E^\dagger_{s,ij,\epsilon} (R_1)
E^\dagger_{s,kl,\epsilon^\prime}(R_2)
E_{s,ij,\epsilon}(R_2)
E_{s,kl,\epsilon^\prime}(R_1)
}.
\label{eq:seconordG00}
\end{equation}

To compare these quantities with the experimental results, we simply select all the relevant transitions contributing to the detected fields, for example, those corresponding to the polarization allowed to reach a particular detector.

Figure \ref{fig:coincidencesP1} shows measured versus predicted
normalized coincidences as a function of the input power for both the
$776\,\mathrm{nm}$ laser and the $780\,\mathrm{nm}$ laser.
The same normalization factor was used for all the
experimental curves as well as for the theoretical calculations.
Both combinations of possible output polarizations are shown;
$G^2_{HV}$ denotes the second-order
correlation between an H-polarized photon emitted from the
$762\,\mathrm{nm}$ transition
and a V-polarized photon from the $795\,\mathrm{nm}$ transition.
Conversely, $G^2_{VH}$ is the second-order
correlation between a V-polarized photon emitted from the
$762\,\mathrm{nm}$ transition
and an H-polarized photon from the $795\,\mathrm{nm}$ transition.
In general, the agreement of our model for
$G_{VH}^2$ with the corresponding experimental
results is not as good as in the case of the counts. However, $G_{VH}^2$
calculations match the experimental data quite reasonably.
When varying the input power of the $776\,\mathrm{nm}$ laser,
agreement is good for low powers of the $780\,\mathrm{nm}$ laser.
This is consistent with the results of the counts section.
When varying the power of the $780\,\mathrm{nm}$ laser,
agreement with the experimental data is also better at lower
input powers for the $776\,\mathrm{nm}$ laser.
At higher powers, the agreement is not as good
as when modeling counts. Both model and experiment expect much lower coincidence rates for the polarization combination VH than for the HV combination. While both polarization combinations are in principle compatible with the usual FWM phase-matching condition, the VH combination seems to be well described by a model that correctly predicts field intensities at the detectors while the HV combination, which is favored by the FWM setup, involves more complex processes, possibly signaling a dependence on collective effects.

\section{Discussion}
\label{sec:discussion}

As shown in Fig. \ref{fig:countsP1}, experiment and theory agree very well for both ports of polarization, $H$ and $V$, at the two photon detection stations. For signal photons ($762\,\mathrm{nm}$), both data and calculations yield a higher number of counts for the $H$-polarized component than for the V-component, roughly by a factor of 2; see panels (a) and (c). This difference is much less pronounced but still present for idler photons ($795\,\mathrm{nm}$), as depicted in panels (b) and (d). For both the expected and measured counts, these differences persist across both input-power sweeps, as can be noted from panels (e) and (g) and (f) and (h). Generally speaking, the model tends to be more accurate at higher pump powers for predicting photon counts, yet works in the full range. 

Coincidences, however, were only modeled correctly for one of the two analyzed intensity correlations. Fortunately, this highlights the ultimate limits of the one-atom model. Recall the curves in panels (c) and (d) of Fig. \ref{fig:coincidencesP1} for $G^2_{VH}$. Both of them resemble the curves for the counts in panels (c) and (g) in Fig. \ref{fig:countsP1}, with reduced amplitudes. This suggests that this correlation behaves similarly to two independent sources. In other words, the probability of measuring a signal photon with polarization V followed by an H-polarized idler photon is governed by the product of the probability of each decay. For $G^2_{HV}$, the behavior is starkly different. The experimental data in panels (a) and (b) in Fig. \ref{fig:coincidencesP1} have a tendency that disagrees qualitatively with the theoretical curves, which also clearly underestimate the data. It is both experimentally and theoretically expected for $G^2_{HV}$ to be larger than $G^2_{VH}$ since the H-component presents a higher number of counts than the respective V-component, Fig. \ref{fig:countsP1} [panels (a) and (c)]. The theoretical curves roughly maintain the 2-to-1 ratio between H-polarized and V-polarized photons present in the photon counts whilst experimental data show nearly a 10-to-1 ratio for the $P_1$ sweep and a 5-to-1 ratio for the $P_2$ sweep.

One possible source for this discrepancy between the measured intensity correlations and the intensity correlations predicted by the single-atom model is the presence of collective effects \cite{Yanes-Thomas:2025}, which modify the statistics of the generated photons through enhanced interatomic interactions between the dipoles that are induced by near-resonance light inside the atomic cloud. These effects lead to atoms altering the overall dynamics of their neighbors during the FWM process. This occurs through two different mechanisms: first, through the coherent exchange of
photons between atoms, which can modify their collective dynamics and emission properties; and second, through incoherent collective radiative processes, such as superradiant
or subradiant emission, which alter the effective decay channels and photon statistics. Such effects could lead to a self-seeding mechanism that is not captured by our single-atom model and would have to be introduced phenomenologically. This effect would also be stronger for the more common decay pathway involving $H$-polarized signal photons.

\section{Conclusions}
\label{sec:conclusion}

We explored the prediction limits of the photon-pair statistics generated by an ensemble of atoms subject to a FWM process, using an exhaustive single-atom model. The system under consideration was a cold cloud of $^{87}\mathrm{Rb}$ atoms confined by a MOT and subjected to a system of transitions with a diamond configuration. In order to capture the full atomic process, not only the FWM lasers but also the re-pump laser of the MOT had to be taken into account. Furthermore, we included each Zeeman level of the six hyperfine levels to enable the prediction of the polarization of the generated photons. Our model thus includes 32 atomic levels and all the relevant experimental parameters: pump and re-pump laser powers and detunings, and the magnetic-field gradient of the MOT coils. Within this framework, we calculate not only the emission rates of the different optical transitions but also the polarization of the emitted photons, allowing a direct comparison with polarization-resolved measurements.

With this model, we were able to directly calculate the first- and second-order coherence functions and compare them to their experimental counterparts, counts and coincidences. In the case of the former, good quantitative agreement over a broad range of laser powers was found between experiment and theory. This demonstrates that an accurate description of the internal atomic structure is essential for modeling the emitted intensities, and that including all relevant Zeeman sub-levels is instrumental for predicting the intensities of the different polarization components of the generated photon pairs. 

Although the agreement with the measured second-order correlation functions is significantly improved with respect to
simplified few-level models, the calculated coincidence rates
systematically underestimate the experimental values for the $G^2_{HV}$ correlation functions. However, they perform better when reproducing the $G^2_{VH}$ correlation functions, which present lower values by at least a factor of 5. This suggests that the former measured coincidences contain coherence from a source other than a single emitting atom, whereas the latter do not. 

Collective effects, such as those emerging from dipole-dipole interactions \cite{Yanes-Thomas:2025, Lehmberg:1970}, can strongly alter atomic dynamics and decay rates and thus could help explain these discrepancies. Propagation effects of light through the atomic cloud are another possible explanation \cite{Leszczynski2017}.

The single-atom model has been a workhorse in atomic physics for decades, even when the system under study is an ensemble. The number of levels has been kept low during its implementation to reduce computational demands and to avoid cumbersome analytical equations. However, with current personal-computer technology and available software, it is possible to change this approach to complete single-atom models. Thus, it is possible to take further advantage of this model by testing the multiple systems being researched in laboratories worldwide. We chose the case of FWM for generating photon pairs correlated in polarization, since it is of interest to us. Nevertheless, this could be just as easily applied to quantum memories \cite{SHINBROUGH2023297,Lei2023}, Rydberg atoms \cite{adams2019}, and optical lattices \cite{lin2025}, just to mention a few examples. Therefore, there is still ample space to work with this model before facing the challenging theoretical task of including collective effects.    

\section{Acknowledgments}
We thank for the support to our laboratory from UNAM through the grants DGAPA-PAPIIT no. IN112624 and CIC-LANMAC 2026, and from SECIHTI via the grants CF-2023-G-15 in the program Frontiers of Science and LNC-2023-51 in the program of National Laboratories. P.Y.-T. was financially supported by CONAHCYT through the Estancias Posdoctorales por México program (2022, Grant No. 3). A.K. was financially supported by the Departamento de Ciencias Básicas, UAM-A, under Grants No. 2232218 and CB003-25 (Grant No. 22322036). We express our gratitude to Carlos A. Gardea-Flores and Rodrigo A. Gutierrez-Arenas for their contributions in the design, construction and implementation of several home-made pieces of instrumentation.

\bibliographystyle{unsrt} 
\bibliography{refs} 

\end{document}